\documentclass[prc,nofootinbib,preprint,tightenlines,amsmath,amssymb ]{revtex4}
\usepackage{graphicx,bm}

\newcommand{\svec}[1]{\bm{#1}}

\begin{document}
\bibliographystyle{apsrev}


\title{Intrinsic-Density Functionals}
\author{J. Engel}
\email[]{engelj@physics.unc.edu}
\affiliation{Department of Physics and Astronomy, CB3255,
             University of North Carolina, Chapel Hill, NC  27599-3255}

\date{October 10, 2006}
\begin{abstract}

The Hohenberg-Kohn theorem and Kohn-Sham procedure are extended to functionals of the
localized intrinsic density of a self-bound system such as a
nucleus.  After defining the intrinsic-density functional, we modify
the usual Kohn-Sham procedure slightly to evaluate the mean-field
approximation to the functional, and carefully describe the construction of the
leading corrections for a system of fermions in one dimension with a
spin-degeneracy equal to the number of particles $N$.  Despite the
fact that the corrections are complicated and nonlocal, we are able
to construct a local Skyrme-like intrinsic-density functional that,
while different from the exact functional, shares with it a minimum
value equal to the exact ground-state energy at the exact
ground-state intrinsic density, to next-to-leading order in $1/N$.
We briefly discuss implications for real Skyrme functionals.

\end{abstract}
%
%
\maketitle
%
%

\section{Introduction}

The Hohenberg-Kohn theorem \cite{Hoh64} and the associated Kohn-Sham
procedure \cite{Koh65} have helped to shift the interpretation of
Skyrme-mean-field theory in heavy nuclei. Nowadays, instead of
focusing on a Skyrme nucleon-nucleon interaction, theorists start
with a Skyrme (or relativistic) mean-field-like density functional,
modifying its parameters to fit nuclear properties in ways that are
not obviously consistent with the existence of an underlying
effective two-body interaction. The Hohenberg-Kohn theorem is often
used to justify this step, the idea being that Skyrme functionals
are an approximation to a universal Kohn-Sham density functional
--- written in terms of single-particle orbits and leading to
Hartree-like equations --- that is guaranteed to include all
correlations.

Because the nucleus is a self-bound system, however, the usual
Hohenberg-Kohn theorem is not really a justification.  It states
that there exists a universal functional of a system's
\emph{laboratory} density that has a minimum at the ground state
density, and that the value of the functional at the minimum is the
ground-state energy.  The adjective ``universal'' means that the
functional can be written as a sum of a part that is independent of
whatever external one-body potential the system is in
--- the universal part
--- and a simple term $\int V(\svec{r}) \rho(\svec{r}) d\svec{r}$,
where $V$ is the one-body potential and $\rho$ is the usual
laboratory density. The density of interest in nuclear physics,
however, is the \emph{intrinsic} density, i.e.\ the density relative
to the nuclear center of mass, not relative to a fixed point in the
lab.  It is this intrinsic density that is localized and measured in
scattering experiments. By contrast, the laboratory nuclear density
of an isolated nucleus is spread out evenly over all space because
the nuclear center of mass is free to move anywhere. Thus, the
universal Hohenberg-Kohn functional has a minimum at a constant
spread-out density that contains no information about the localized
intrinsic density.  The guaranteed existence of this functional is
irrelevant to nuclear physics.  The same is true of the usual
Kohn-Sham procedure, which represents the density as a sum of
squares of single-particle wave functions, leading to a Hartree-like
equations for the ground-state density and energy.  The
single-particle orbitals for the ground-state laboratory density are
plane waves.

Because the ground state wave function of the nucleus factors into
center-of-mass and intrinsic parts, it is possible --- by removing the
center-of-mass kinetic energy from the Hamiltonian and adding a potential such
as $V_{CM}\equiv (\sum_i \svec{r}_i)^2$ that is minimized when the center of
mass is at the origin --- to construct a related Hamiltonian for which the
ground-state  laboratory density is the same as the intrinsic density.  One
could then try to produce an ordinary density functional for this new
Hamiltonian, assured by the Hohenberg-Kohn theorem that its minimum would yield
the true intrinsic density.  By making the potential very weak, one could even
make the functional yield the true ground-state energy when minimized.  This
procedure, however, would single out one point in space and would therefore
look very different from Skyrme mean-field theory, in which the functional is
minimized by a family of Slater determinants, related to one another by
translation. Furthermore, the local-density approximation (LDA) and extensions,
through which Skyrme functionals might be derived \cite{Neg70}, could easily
destroy the conditions that force the intrinsic density to be the same as the
laboratory density.  The whole Skyrme enterprise would make much more sense if we knew
that we could construct an exact Kohn-Sham
functional of of the intrinsic density rather than the laboratory density,
without modifying the Hamiltonian.

These considerations, and related ones connected with deformation
and rotation, lead us to ask whether analogs of the Hohenberg-Kohn
theorem and Kohn-Sham procedure exist for the intrinsic density.
Rotation is more complicated than translation because it doesn't
decouple from internal motion and because the intrinsic rotational
density is not directly observable, so for now we worry only about
translation and define the intrinsic density simply as the density
with respect to the center of mass. The operator corresponding to
that quantity is
\begin{equation}
\label{eq:def1} \hat{\rho}_I(\svec{r}) \equiv
\hat{\rho}(\svec{r}+\hat{\svec{r}}_{\rm CM})~,
\end{equation}
where $\hat{\rho}$ is the usual laboratory density operator, the
center-of-mass position operator $\hat{r}_{\rm CM}$ is defined by
\begin{equation}
\label{eq:def2}
\hat{\svec{r}}_{\rm CM} \equiv \frac{1}{N}\sum_{i=1}^N \hat{\svec{r}}_i~,
\end{equation}
and the hats over $\rho_I$, $\rho$, $\svec{r}_i$, and $\svec{r}_{\rm
CM}$ are to emphasize that the quantities are operators, unlike the
variable $\svec{r}$ in Eq.\ (\ref{eq:def1}).  The intrinsic density,
unlike its lab-frame counterpart, is an $N$-body operator,
complicating density-functional theory for self-bound systems.

In the next section we address the question of whether there is an
analog of the Hohenberg-Kohn theorem guaranteeing the existence of
an intrinsic ``universal" intrinsic-density functional with a
minimum at the exact ground-state intrinsic density and energy.
Section \ref{sec:ks} argues that a Kohn-Sham-like procedure can be
used to find the exact intrinsic density, and outlines the
construction of the corresponding Kohn-Sham equations.  We will work
with a specific one-dimensional model to simplify the formalism, but
will carry the construction as far as possible; as a result this
section is the longest and most technical.  In section \ref{sec:ph}
we produce a phenomenological Skyrme-like functional that, while not
exact away from the minimum, does yield the model's exact intrinsic
density and energy at its minimum to good accuracy.  We then briefly
discuss the implications of these results for real three-dimensional
nuclear physics.

\section{Hohenberg-Kohn Theorem for Intrinsic Density}

There is indeed a Hohenberg-Kohn theorem for the intrinsic density.
In fact, Valiev and Fernando \cite{Val97} have shown --- assuming a
nonrelativistic Hamiltonian, and using an effective-action formalism
invented in Ref.\ \cite{Fuk94} --- that given \emph{any} Hermitian
operator $\hat{Q}(\svec{r})$, one can construct an energy functional
$E[Q]$ that is universal in a sense to be discussed and has a unique
(and correct) minimum at $Q(\svec{r})=\langle
\hat{Q}(\svec{r})\rangle_{\rm gs}$, the ground-state expectation
value of $\hat{Q}(\svec{r})$. Introducing a ``source'' $J(\svec{r})$
that couples to $Q(\svec{r})$ and alters the ground-state energy:
\begin{equation}
\label{eq:calE}
{\cal E}[J]\equiv \langle \hat{H} +\int
J(\svec{r}) \hat{Q}(\svec{r}) d\svec{r} \rangle_{\rm gs}^J~,
\end{equation}
one can define the functional $E[Q]$ as the functional Legendre
transform with respect to $J$. In other words
\begin{equation}
\label{eq:denf}
E[Q] = {\cal E}[J[Q]] - J[Q] \circ Q\quad,
\end{equation}
where $J[Q](\svec{r})$ is the particular source function that makes
$\langle \hat{Q}(\svec{r}) \rangle_{\rm gs}^J = Q(\svec{r})$, and we have used
the convention $A \circ B \equiv \int A(\svec{r}) B(\svec{r})
d\svec{r}$. It's not hard to show \cite{Wei96} (and this is an
alternative and perhaps clearer definition) that $E[Q]$ is the
minimum of $\langle \hat{H} \rangle$ over all normalized
$N$-particle wave functions that have $\langle
\hat{Q}(\svec{r})\rangle = Q(\svec{r})$.   Since $\langle \hat{H}
\rangle$ is smallest in the ground state, the functional is
minimized when $Q \equiv \langle \hat{Q} \rangle_{\rm gs}$.  It is
universal in that the addition of an arbitrary term $ V \circ
\hat{Q}$ to the Hamiltonian merely adds $V \circ Q$ to $E[Q]$.

All this means, in particular, that for nuclei there is a universal
``intrinsic-density functional'' $E[\rho_I]$ with a minimum at the
ground-state intrinsic density, and that we know in principal how to
construct it.  It's worth noting that the functional is not
universal under the addition of an ordinary one-body potential
$\sum_i V(\hat{\svec{r}}_i)$ to the Hamiltonian, but rather the
addition of a generically $N$-body potential $\sum_i
V(\hat{\svec{r}}_i-\hat{\svec{r}}_{\rm CM})$ that affects only intrinsic
structure.

The important question, though, is how to incorporate orbitals into this
framework in a way that leads to a constructive procedure for the
energy functional.  In other words, can one derive a Kohn-Sham-like (e.g.\
Skyrme-like) functional for the intrinsic density, even in principle?  In the
next section we show that one can.

\section{Orbitals and Kohn-Sham-like equations \label{sec:ks}}

\subsection{Framework}

Refs.\ \cite{Val97,Pug02} introduce Kohn-Sham orbitals in
functionals of the ordinary density (in systems that are not self
bound) through the perturbative ``inversion method'' \cite{Oka96}.
In this approach one divides $\tilde{J}[\rho]$, where the tilde here
and below distinguishes the quantity from the related one to be used
here for the intrinsic density $\rho_I$, into a piece
$\tilde{J}_0[\rho]$, the Kohn-Sham potential that forces a
\emph{noninteracting} system to have the density $\rho$, and the
rest.  The entire functional can be constructed from $\tilde{J}_0$
and expressed in terms of the noninteracting orbitals $\phi_k$ and
energies $\epsilon_k$ that $\tilde{J}_0$ determines. In our
notation, under the assumption that $\tilde{\cal E}[\tilde{J}]$ can
be divided into successively smaller contributions $\tilde{\cal
E}[\tilde{J}]=\tilde{\cal E}_0[\tilde{J}] + \tilde{\cal
E}_1[\tilde{J}] + \cdots$, the energy-density functional
$\tilde{E}[\rho]$,  has the form \cite{Val97,Pug02}
\begin{equation}
\label{eq:inver}
\tilde{E}[\rho]= \tilde{E}_0[\rho] + \tilde{E}_{\rm
int}[\rho]~,
\end{equation}
where
\begin{equation}
\label{eq:inver1}
\tilde{E}_0[\rho]=\tilde{\cal E}_0[\tilde{J}_0] -  \tilde{J}_0
\circ \rho~, \qquad \tilde{E}_\textrm{int}[\rho]=
\tilde{\cal E}_1[\tilde{J}_0]+\cdots~,
\end{equation}
and we have truncated $\tilde{E}_\textrm{int}[\rho]$ at lowest order.
To find the ground-state density and energy one rewrites the minimization condition
$\delta \tilde{E}[\rho]/\delta\rho = 0$, as an implicit Kohn-Sham equation of the form
\begin{equation}
\frac{\delta \tilde{E}_{\rm int}[\rho]}{\delta \rho} = \tilde{J}_0[\rho]~,
\end{equation}
where we have used Eq.\ (\ref{eq:inver}) and the fact that $\delta
\tilde{E}_0/\delta[\rho]=-\tilde{J}_0$, which follows from the
definition of $\tilde{E}_0$ as a Legendre transform. The Kohn-Sham
procedure then consists of choosing a trial potential $\tilde{J}_0$,
solving the single-particle Schroedinger equation to obtain the
corresponding $\phi_k$ and $\epsilon_k$, computing $\tilde{E}_{\rm
int}$ and $\delta \tilde{E}_{\rm int}/\delta\rho$ in terms of these
quantities, and then iterating until the energy and density
converge.  The procedure amounts to solving Hartree equations with
the mean-field potential $V[\rho]=\delta \tilde{E}_{\rm int}/\delta
\rho$.

We can apply this procedure to our intrinsic functional $E[\rho_I]$,
even though $\rho_I$ is an $N$-body operator, provided we start at a
slightly different point:  mean-field theory.  To make the
discussion less abstract, we refer explicitly to a simple
one-dimensional system. Before constructing an energy functional, we
describe the system, its exact ground-state energy and intrinsic
density, and the mean-field approximations to these quantities.

\subsection{Model in one dimension}

We consider $N$ fermions with spin-degeneracy $N$ interacting
via an attractive two-body delta function ($\hbar=1$):
\begin{equation}
H=-\frac{1}{2m}\sum_{i=1}^N \frac{d^2}{dx_i^2} - g \sum_{i<j}
\delta(x_i-x_j)\quad.
\end{equation}
This model has the nice feature of being exactly solvable.   The
ground state is bound and the corresponding energy and intrinsic
density are known \cite{Mcg64,Cal75}:
\begin{eqnarray}
E_{\rm gs}&=&\frac{1}{24} m g^2(N^3 - N) \label{eq:exacte} \\
\rho_I^{\rm gs}(x)&=& mg \sum_{n=1}^{N-1}(-)^{n+1}\frac{n(N!)^2
e^{-nNmg|x|}}{(N+n-1)!(N-n-1)!}\label{eq:exactden}\\
&=&\frac{mgN^2}{4 \textrm{cosh}^2(\frac{Nmgx}{2})}
\left(1-\frac{1}{N}\left[1-\frac{3}{2 \textrm{cosh}^2(\frac{Nmgx}{2})}\right]
+{\cal
O}(\frac{1}{N^2} )\right)~, \qquad mg|x| \gtrsim \frac{3 \, {\rm ln}N}{N^2} ~.\nonumber
\end{eqnarray}
The ground state differs from that of a nucleus, however, in that
the system shrinks to a point as $N$ increases and the binding
energy goes like $N^3$; there is nothing resembling saturation.  A
local-density approximation for such a system, which has no uniform
limit, is hard to imagine.

Ref.\ \cite{Cal75} analyzed the Hartree-Fock approximation, to which
we refer as the mean-field approximation, for the ground state. The
approximation is improved when the center-of-mass kinetic energy is
subtracted from the Hamiltonian, as e.g.\ in the no-core shell model
\cite{Nav00}.  Subtraction makes a localized ground-state possible,
since the center-of-mass is no longer required to be in a plane-wave
state. Because the spin degeneracy is equal to the number of
particles, whether or not one subtracts, the spatial mean-field
orbital $\phi(x)$ is the same for every particle. The mean-field
equation for $\phi$, with no external source and the center-of-mass
kinetic energy subtracted, is
\begin{equation}
\label{eq:hfeqn}
-\frac{N-1}{2mN}\phi^{\prime\prime}(x)-(N-1)g|\phi(x)|^2 \phi(x) = \epsilon
\phi(x)
\end{equation}
Because of translational invariance, the solutions make up a degenerate set with each
member centered at a different point
$x=a$ in space:
\begin{equation}
\label{eq:sol}
\phi(x)= \frac{\sqrt{Nmg}}{2 \textrm{cosh}[Nmg(x-a)/2]}~, \qquad
\epsilon=-\frac{mg^2}{8}N(N-1)~.
\end{equation}
The average position $\bar{x}$, given by
\begin{equation}
\bar{x}=\int x|\phi(x)|^2 dx~,
\end{equation}
is just $\bar{x}=a$ for the mean-field solutions.
The mean-field energy is then
\begin{eqnarray}
\label{eq:emf}
E_{\rm mf} &=& N(N-1) \int dx \left[ \frac{1}{2mN}|\phi^{\prime}(x)|^2
-\frac{g}{2}|\phi(x)|^4 \right] \nonumber \\
&=& N \epsilon + \frac{N(N-1)g}{2}\int |\phi(x)|^4 dx
\nonumber\\
& =& -\frac{mg^2}{24}N^2(N-1)
\end{eqnarray}

Interestingly, the mean-field energy is correct at leading order in
the small quantity $1/N$, that is, the ${\cal O}(N^3)$ term is
right. At ${\cal O}(N^2)$ there is an error; the exact energy in
Eq.\ (\ref{eq:exacte})has no term of that order.  The mean-field
${\cal O}(N^2)$ energy is reduced by the subtraction of the
center-of-mass kinetic energy, but, as Ref.\ \cite{Yoo77} shows, one
needs to include ring diagrams\footnote{One might mistakenly infer
from Ref.\ \cite{Yoo77} that ring diagrams with more than two rings
contribute very little to the ${\cal O}(N^2)$ correlation energy.
That can appear to be so, but only if one modifies the Hamiltonian
(by subtracting the center-of-mass energy) in the Hartree-Fock
calculation but neglects to do so in the RPA.} (i.e.\ the RPA
correlation energy) to cancel it completely.

The mean-field laboratory density $N\phi^2(x)$ is also correct in
some sense in leading order; because center-of-mass motion becomes
irrelevant at large $N$, the mean-field laboratory density
reproduces the exact intrinsic density there, assuming the center of
mass is at the origin \cite{Cal75}. The mean-field \emph{intrinsic}
density can be computed by expanding the operator $\hat{\rho}_I(x)
\equiv \hat{\rho}(x+\bar{x}+\hat{x}_{\rm CM}-\bar{x})$ around
$x+\bar{x}$ in powers of $\hat{x}_{\rm CM}-\bar{x}$, leading, after
some algebra, to
\begin{eqnarray}
\label{eq:intrsd}
\langle  \hat{\rho}_I(x) \rangle_{\rm mf}
&=&c_NN\left[|\phi(c_N x+\bar{x})|^2 + \frac{\mu}{2N}
\frac{d^2}{dx^2}|\phi(c_N x+\bar{x})|^2+{\cal O}(\frac{1}{N^2}) \right] \\
& =& N  \left[
|\phi(x+\bar{x})|^2 + \frac{1}{N}\left(|\phi(x+\bar{x})|^2 + x
\frac{d}{dx}|\phi(x+\bar{x})|^2 +  \frac{\mu}{2}
\frac{d^2}{dx^2}|\phi(x+\bar{x})|^2\right)  +{\cal O}(\frac{1}{N^2}) \right] ~,
\nonumber
\end{eqnarray}
with
\begin{equation}
\label{eq:def_mu_cn}
\mu=\int(x-\bar{x})^2 |\phi(x)|^2 dx \quad, \qquad c_N=\frac{N}{N-1}
\end{equation}
Beyond leading order, this expression gives a better approximation
to the exact intrinsic density than does the mean-field laboratory
density $N|\phi(x)|^2$, as can be seen, for example, by using it to
compute the $\langle x^2 \rangle_{\rm intr}$.  The result,  $\langle
x^2 \rangle_{\rm intr} = \mu (1-\frac{1}{N})$, is the same as the
exact one.

With these results in hand, we can move to the construction of a
Kohn-Sham-like intrinsic-density functional for the model.

\subsection{Intrinsic Density Functional at Mean-Field Level}

The problem with applying the Kohn-Sham procedure directly to the
intrinsic density is that the source term, $J_0 \circ \hat{\rho}_I$,
is an $N$-body operator.  We cannot treat the source exactly and
still have a noninteracting system, the starting point for the
inversion method discussed above.  Instead, we start with the
mean-field approximation, not just for the true ground-state, but
also for the ground state with the addition of the term $J_0 \circ
\hat{\rho}_I$.  In other words, $J_0[\rho_I]$ is now the source
function that forces the intrinsic density to be $\rho_I$ when the
system is treated in \emph{mean-field approximation}. The resulting
mean-field single-particle orbitals will play the same role in
obtaining the exact intrinsic functional as Kohn-Sham orbitals for a
noninteracting system do in the usual approach, and the inversion
method will go through without any other modifications.  Of course,
to apply it we must assume that corrections to mean-field theory are
perturbative.  For the model just described that is certainly the
case, as we shall see.  For real Skyrme functionals, it is not
unreasonable to assume the existence of an effective Hamiltonian for
which Hartree-Fock is a good starting point, at least in spherical
nuclei.

Constructing the mean-field intrinsic-density \emph{functional} as
the first term $E_0[\rho_I]$ in the intrinsic version of Eq.\
(\ref{eq:inver}) can be done without solving any equations.  We are
supposed to start with
\begin{equation}
\label{eq:wmf} {\cal E}_0[J_0]= N(N-1) \int dx \left[
\frac{1}{2mN}|\phi^{\prime}(x)|^2 -\frac{g}{2}|\phi(x)|^4
\right] + J_0 \circ \langle \hat{\rho}_I \rangle_{\rm mf}~,
\end{equation}
where $\langle \hat{\rho}_I \rangle_{\rm mf}$ is given by Eq.\
(\ref{eq:intrsd}) and we have taken the useful but not essential
step of subtracting the center-of-mass kinetic energy (though
without explicitly writing a more complicated two-body
center-of-mass term that has no effect on bound states)
to improve the mean-field approximation.  But since $J_0$ is
supposed to be fixed so that $\langle \hat{\rho}_I \rangle_{\rm mf}
= \rho_I$ and then $J_0 \circ \rho_I$ subtracted from ${\cal
E}_0[J_0]$, the mean-field functional
looks just like the first line of Eq.\ (\ref{eq:emf}):
\begin{equation}
\label{eq:e0}
E_0[\rho_I]=
 N(N-1) \int dx \left[ \frac{1}{2mN}|\phi^{\prime}(x)|^2
-\frac{g}{2}|\phi(x)|^4 \right] 
\end{equation}
This simple Skyrme-like object, when written in terms of the orbital
$\phi$, is the same as the mean-field functional one would get for
the ordinary one-body density, but now \emph{$\phi$ is a functional
of $\rho_I$} (up to the location of the center of mass $\bar{x}$,
which must also be specified) rather than $\rho$.  The relation
between the two functionals is not surprising because at mean-field
level the Slater determinant constructed from orbital $\phi$ is the
system's wave function, and the energy functional --- the energy
corresponding to that wave function --- is just the expectation
value of the Hamiltonian, no matter what observable we write the
functional (and $\phi$) in terms of. (When we go beyond mean-field
theory, center-of-mass motion makes the functionals of the lab and
intrinsic density quite different.)

In more detail, $\phi$ is a functional of $\rho_I$ in the following
sense: $\rho_I$ determines $J_0$ through the solution to $\delta
{\cal E}_0/\delta J_0=\rho_I$ (Ref.\ \cite{Val97} shows that the
solution exists and is unique) and $J_0$ determines $\phi$, again up
to its overall location, through the solution to the mean-field
equations, which we haven't yet written, that come from varying
${\cal E}_0[J_0]$ with respect to $\phi$ (with a constraint on its
norm).  With $\phi$ and $E_0$ functionals of $\rho_I$, one can then
reproduce the mean-field ground-state energy and intrinsic density
by solving the equation $\delta E_0/\delta \rho_I=0$.  Because this
equation says that $J_0=0$, finding the minimum corresponds to
solving the source-free mean-field equation, {\it i.e.}\ the Hartree
equation Eq.\ (\ref{eq:hfeqn}), and then computing $\rho_I$ from
Eq.\ (\ref{eq:intrsd}).   This Hartree equation is the leading
approximation to the Kohn-Sham-like equation we seek.

All this may seem pedantic, but it contains an important point about
the relation of $\phi$ to $\rho_I$:  the condition $\delta {\cal
E}_0/\delta J_0=\rho_I$ simply states that $\rho_I$ will be given in
terms of $\phi$ by Eq.\ (\ref{eq:intrsd}), with $\rho_I$ replacing
$\langle \hat{\rho}_I\rangle_{\rm mf}$.  Thus, though $\rho_I(x)$ is
just $N|\phi(x+\bar{x})|^2$ to leading order in $1/N$, it has
higher-order corrections. This fact is apparently important for real
Skyrme functionals, which fit data best when a prescription
\cite{Sch91} for $\rho_I$ that approximates the real-world analog of
the Eq.\ (\ref{eq:intrsd}) is used to calculate observables such as
the rms radius \cite{Rei06}. Using the uncorrected expression
$\rho_I(x) \equiv \sum_k |\phi_k(x+\bar{x})|^2$ for this purpose
produces noticeably poorer results.

It actually is possible to formulate an intrinsic-density functional
theory in which $\rho_I \equiv \sum_k |\phi_k(x+\bar{x})|^2$ without
corrections. In the simple model we can do so by neglecting the
${\cal O}(1/N)$ terms in Eq.\ (\ref{eq:intrsd}) when defining ${\cal
E}_0[J_0]$, and taking their effects into account at the next level
of approximation.  Such a step would in fact simplify much of the
formalism to come, e.g., the mean-field equation for the orbital
$\phi$ in the presence of the many-body source $J_0$.  And when
expressing the leading-order intrinsic-density functional in terms
of $\phi$ we'd still end up with Eq.\ (\ref{eq:e0}).   That object,
however, would be a different functional of $\rho_I$ than before and
would produce poorer results for the ground-state intrinsic density.
We will not elaborate the corrections to the functional necessary to
restore the lost accuracy, lest this paper get even longer than it
is.  Instead, using the relation between $\rho_I$ and $\phi$ given
by Eq.\ (\ref{eq:intrsd}), we proceed to take up corrections that
are genuinely beyond mean-field theory.  Though the lack of
something like the LDA prevents us from actually calculating the
corrections, either analytically or numerically, we describe how to
do the latter, in some detail.

\subsection{Beyond Mean-Field Theory}

To go beyond the relatively straightforward mean-field approximation
to $E[\rho_I]$, one would need to calculate $E_{\rm int}[\rho_I] =
E_1[\rho_I] + \cdots = {\cal E}_1[J_0] + \ldots $. (We content
ourselves with $E_1$ here, though we could go further.) Then,
provided one could calculate $\delta E_1/\delta \rho_I(x)$, one
could follow the procedure Kohn-Sham procedure outlined above,
rewriting the minimization condition $\delta E[\rho_I]/\delta \rho_I$
in the form
\begin{equation}
\label{eq:kseq}
J_0(x) =
\frac{\delta{\cal  E}_1[J_0]}{\delta \rho_I(x)} ~.
\end{equation}
and solving by iteration, starting with a guess for $J_0$.  We will call Eq.\
(\ref{eq:kseq}) the Kohn-Sham equation, even though strictly speaking it doesn't
involve Kohn-Sham orbitals.

Given a $J_0$, one could find ${\cal E}_1$ in the following way:
The first step would be to find the associated mean-field orbital $\phi$ by
solving the mean-field equation, which we finally have to write
down, for the system in the source $J_0 \circ \hat{\rho}_I$.  We obtain the
equation
by varying $\phi$ in Eq.\ (\ref{eq:wmf}) and using Eq.\ (\ref{eq:intrsd}).
Since $\bar{x}$ in the latter
is itself a functional of $\phi$, the result
is complicated:
\begin{eqnarray}
\label{eq:mfeqj}
-\frac{\phi^{\prime\prime}(x)}{2mc_N}-\left[(N-1)g|\phi(x)|^2
-J_0^{\bar{x}}(x)                        
+\frac{(x-\bar{x})}{c_N} \int
J_0^{\bar{x}}{}^{\prime}(y)              
|\phi(y)|^2 dy \right] \phi(x) \hspace{3.3cm}&&  \\
+ \frac{\mu}{2 N} \left[
J_0^{\bar{x}}{}^{\prime\prime}(x)        
-(x-\bar{x})\int J_0^{\bar{x}}{}^{\prime\prime\prime}(y)
|\phi(y)|^2 dy + \frac{(x-\bar{x})^2}{\mu}
\int J_0^{\bar{x}}{}^{\prime\prime}(y)   
|\phi(y)|^2 dy \right] \phi(x) = \ \epsilon \phi(x)~,&& \nonumber
\end{eqnarray}
where $J_0^{\bar{x}}(x) \equiv J_0(\frac{x-\bar{x}}{c_N})$ and we have been
(and will continue to be) cavalier about factors of $c_N$ in terms of ${\cal
O}(1/N)$, since they make a difference only in terms of ${\cal O}(1/N^2)$, which we
have dropped. They could be included in a more accurate functional.
We have
also added $\bar{x}$'s in the fourth and sixth terms to keep
translational invariance manifest; they would otherwise be absorbed
into $\epsilon$.

Next, one would need an approximation in which to evaluate/define
${\cal E}_1$. As mentioned above, Ref.\ \cite{Yoo77} demonstrates
that in the absence of a source the largest corrections to the
mean-field approximation in a $1/N$ expansion come from summing ring
diagrams. The same statement does not apply to the intrinsic density
functional at all possible densities, but the ring sum, which is the
leading correction to the mean-field in a ``loop-expansion''
\cite{Neg88} of ${\cal E}[J]$, should yield the $1/N$ corrections
near the ground-state intrinsic density and makes a natural choice
for ${\cal E}_1$.

To obtain an expression for the ring sum, we work with the one-body
density matrix $\rho_{ab}$ \cite{Rin80}. We can use it, for
starters, to represent the contents of Eq.\ (\ref{eq:mfeqj}). The
left-hand side of that equation, with a delta function
$\delta(x-x')$ replacing $\phi(x)$, gives the coordinate-space
representation of the mean field $h(x,x')[J_0]$ . We can write this
$h$ (with spin now
included) in an arbitrary space+spin
basis in terms of $\rho_{ab}$\,:
\begin{eqnarray}
\label{eq:hab}
h_{ab}[J_0] &=&\frac{\partial {\cal E}_0[J_0]}{\partial \rho_{ba}} \\
&=&
\cdots +(J_0^{\bar{x}})_{ab}
 -\frac{1}{c_N N}
 {\rm Tr}[ (J_0^{\bar{x}})^{\prime} \rho] \,(x-\bar{x})_{ab}
\nonumber \\
&&+\frac{1}{2 N^2} \left( \mu N(J_0^{\bar{x}})^{\prime\prime}_{ab}
- \mu {\rm
Tr}[(J_0^{\bar{x}})^{\prime\prime\prime}\rho] \, (x-\bar{x})_{ab}
 +{\rm Tr}[(J_0^{\bar{x}})^{\prime\prime}\rho] \ [(x-\bar{x})^2]_{ab}
\right)~,\nonumber
\end{eqnarray}
where Tr represents a trace over matrix elements, and we have only
omitted the terms that do not depend on $J_0$ in the second and
third lines. Hartree-Fock theory corresponds to finding the basis
that make $h$ diagonal.  Again, we have inserted two $\bar{x}$'s to
manifest translational invariance; they just add a constant to all
eigenvalues, without altering the Hartree-Fock basis.

The ring sum, in these terms, is just the RPA correlation energy
calculated with the effective two-body interaction \cite{Rin80}
\begin{equation}
\label{eq:veff}
V_{ab,cd}=\frac{\partial h_{ac}[J_0]}{\partial \rho_{db}} = \frac{\partial^2
{\cal E}[J_0]}{\partial \rho_{ca} \partial \rho_{db}} \quad.
\end{equation}
In coordinate space, this interaction
takes the form 
\begin{eqnarray}
\label{eq:coord}
V(x_1,x_2)=-\frac{g}{c_N}\delta(x_1-x_2)
+ \frac{1}{2mN}\hat{\mathrm{p}}_1 \hat{\mathrm{p}}_2 \hspace{9cm}&& \\
-\frac{1}{N}\left[(x_1-\bar{x}) \
J^{\prime}_0(x_2-\bar{x}) + (x_2-\bar{x})\ J^{\prime}_0(x_1-\bar{x})-  (x_1-\bar{x})(x_2-\bar{x})\left( \int
\! J^{\prime\prime}_0(y-\bar{x})|\phi^2(y)| dy\right) \right]&&\nonumber\\
+\frac{1}{N^2}[ \cdots ]~, \hspace{14.5cm} \nonumber
\end{eqnarray}
where the term with $\hat{\rm p}_i \equiv -i d/dx_i$ is the two-body
center-of-mass Hamiltonian,  matrix elments of that term alone are to be
antisymmetrized, and we have replaced some more
complicated terms with an ellipsis.   The
extra $\bar{x}$'s we added to $h$ (and to Eq.\
(\ref{eq:coord})) have no effect because the particle-hole RPA involves
only terms $h_{ac}$ with $a \neq c$ in Eq.\ (\ref{eq:veff}) and $\bar{x}$ is
just a number, with no off-diagonal matrix elements.

The sum of all the ring diagrams can be expressed as
\cite{Bla86,Shi00}
\begin{equation}
\label{eq:rpa}
E_1[\rho_I] = {\cal E}_1[J_0] = \frac{1}{2\pi}\int_0^{\infty} {\rm
Re}\left({\rm Tr}\left[{\rm ln}(1-\hat{R}(i\omega) \hat{V})+ \hat{R}(i \omega)
\hat{V}\right]\right) \, d\omega \quad,
\end{equation}
where the coordinate-space matrix elements of $\hat{R}(\omega)$ make
up the ``unperturbed'' response function
$R(x_1,x_2;x^\prime_1,x^\prime_2;\omega)$ for the system in the
$J_0$-dependent
mean field. (The spin indices, which we've omitted, just contribute factors
of $N$ in Eq.\ (\ref{eq:rpa}).) For zero-range interactions and our source we need these
matrix elements only at $x_1=x_2 \equiv x$, $x_1^\prime=x_2^\prime
\equiv x^\prime$, and there the
response can be written (for $\phi$ real) as:
\begin{equation}
\label{eq:response}
R(x,x^\prime;\omega) = \phi(x)\phi(x^\prime)\left[\langle x |
\frac{1}{\omega+\epsilon+i\eta-\hat{h}[J_0]} |x^\prime\rangle + \langle x^\prime |
\frac{1}{-\omega+\epsilon +i\eta -\hat{h}[J_0]}|x\rangle \right]~.
\end{equation}
The matrix elements are just single-particle Green's functions for particles
in the mean field.  Equation (\ref{eq:rpa}) is then the RPA
correction to the intrinsic density functional.

One advantage of our definition of intrinsic density (for others, see, e.g.,
Refs.\ \cite{Cal75,Van98}) is that ${\cal
E}[J]$ is translationally invariant; moving $\phi(x)$ doesn't change
it because it doesn't change $\rho_I$, the quantity to which $J$
couples.  As a result, the RPA equations have a zero mode.  With the
center-of-mass kinetic energy subtracted subtracted from the
Hamiltonian, the zero mode contributes nothing to $E_1$, but that
convenience is not an essential part of the treatment. We could
actually start with the full Hamiltonian, with no
center-of-mass-energy subtraction, and recover the same result for
$E_0+E_1$.  The mean-field energy $E_0[\rho_I]$ would be different,
but the difference would be made up \cite{Rin80} by the contribution of the RPA
zero mode to $E_1$, the ring sum in Eq.\ (\ref{eq:rpa}).
Further corrections might require mixing of mean-fields with
different values of $\bar{x}$ to account for center-of-mass motion,
but at RPA order a single-mean field with a proper treatment of the
zero mode is sufficient, even without the trick of removing the
center-of-mass energy from the Hamiltonian.

Despite this nice feature, Eq.\ (\ref{eq:rpa}) is still a
complicated implicit representation of $E_1[\rho_I]$.  To solve the
Kohn-Sham equation, Eq.\ (\ref{eq:kseq}), we need the derivative of
this functional. Taking the derivative is the most numerically
involved step in the entire process\footnote{Ref.\ \cite{Eng97}
discusses some of the difficulties in ``orbital-dependent
functionals'' for atomic and condensed-matter physics.}.  References
\cite{Val97,Pug02} suggest using the relation
\begin{equation}
\label{eq:equal}
\frac{\delta E_1[\rho_I]}{\delta \rho_I(x)} =\frac{\delta {\cal E}_1}{\delta
J_0} \circ \frac{\delta J_0}{\delta \rho_I(x)} \quad,
\end{equation}
In our model the two terms on the right-hand side can be manageably
combined, particularly if we're willing to sacrifice the quality of
the functional away from the minimum, a step that, as we shall see,
has no effect on the Kohn-Sham solution.

To get $\frac{\delta J_0}{\delta \rho_I}$, one can
start by writing $\phi$ in terms of $\rho_I$ and $\bar{x}$ through
the perturbative inversion of Eq.\ (\ref{eq:intrsd}) (assuming
$\phi$ real):
\begin{equation}
\label{eq:phiofrhoi} \phi(x)= \left(c_N
N\right)^{-1/2} \sqrt{\rho_I(\frac{x-\bar{x}}{c_N})
-\frac{\mu}{2N} \rho_I^{\prime\prime}(\frac{x-\bar{x}}{c_N})
+ {\cal O}(\frac{1}{N^2})}~,
\end{equation}
Then one could solve Eq.\ (\ref{eq:mfeqj}) in reverse to obtain
$J_0-\epsilon$ in terms of $\phi$. ($J_0$ is needed only up to a
constant to evaluate higher-order corrections.) Though Eq.\
(\ref{eq:mfeqj}) for $J_0$ is a complicated differo-integral
equation, we can simplify it without harm.   Near the ground-state
value of $\rho_I$, the source $J_0$, which is close to $\delta E_1/\delta \rho_I$ there,
is smaller by ${\cal O}(1/N)$ than $(N-1)g|\phi|^2$, the
$J_0$-independent part of the mean-field equation. $E_1$ is already
smaller than $E_0$ by ${\cal O}(1/N)$ near the ground state, and the
contributions to $E_1$ of terms in which $J_0$ is multiplied by
$1/N$ are therefore down from $E_0$ by ${\cal O}(1/N^2)$. At that
order we're already missing terms because of the restriction of the
relation between $\rho_I$ and $\phi$ in Eq.\ (\ref{eq:intrsd}) to
${\cal O}(1/N)$. Thus, in treating $E_1$ or its derivative we can
neglect the entire second line of Eq.\ (\ref{eq:mfeqj}), (and even
the last term on the first line because the $\phi$ corresponding to
the ground state is symmetric) without changing the Kohn-Sham
solution at ${\cal O}(1/N)$.  This converts the differo-integral
equation for $J_0$ into a simple algebraic one, yielding
\begin{equation}
\label{eq:j0}
J_0^{\bar{x}}(x)-\epsilon \approx \frac{1}{2m}\frac{\phi^{\prime\prime}(x)}{\phi(x)}
+Ng|\phi(x)|^2~.
\end{equation}
With this result and Eq.\ (\ref{eq:phiofrhoi}) it's easy to get 
$\delta J_0(y)/\delta
\rho_I(x)$ through the chain rule; it would vanish for $x \neq y$ with 
the approximations just mentioned.  One could then extract $\delta E_1[\rho_I]/\delta
J_0(y)$ by making local variations in in $J_0(y)$ and calculating the changes
in $E_1[\rho_I] = {\cal E}_1[J_0]$ through Eq.\ (\ref{eq:rpa}).  One
could simplify that equation as well by neglecting all but the
analogs of the first term after the ellipsis in Eq.\ (\ref{eq:hab}) and the first two
lines of the effective interaction, Eq.\ (\ref{eq:veff}), without
affecting the functional near its minimum at next-to-leading order.

That leaves, finally, the Kohn-Sham equation itself.  As mentioned
already, one can solve it iteratively.  For our intrinsic-density
functional that means starting with a guess for $J_0$, solving the
mean-field Eq.\ (\ref{eq:mfeqj}) to obtain $\phi$ and $\epsilon$,
evaluating $E_1$ via Eq.\ (\ref{eq:rpa}) and then $\delta E_1/\delta
\rho_I(x)$ as just discussed, resetting $J_0$ to equal $\delta
E_1/\delta \rho_I$, and repeating until $J_0$, $E_1$, and $\phi$
(and therefore $\rho_I$) converge.  This is equivalent to solving
the mean-field equation  (\ref{eq:mfeqj}) with $J_0$ replaced by
$\delta E_1/\delta{\rho_I}$.  And again, because even the leading
order terms in $E_1$ are down by ${\cal O}(1/N)$ from $E_0$ near the
minimum, we can throw away the higher-order terms in that equation
without affecting the relevant part of $E_1$.  We can even neglect
them in constructing $E_0[J_0[\rho_I]]$ near the ground state,
because in that functional corrections to the mean-field energy,
Eq.\ (\ref{eq:emf}), are \emph{second} order in the small source
$J_0$.   One can get the entire density functional correct at
next-to-leading order near the minimum with a Kohn-Sham equation
that takes simple Hartree form!

For real nuclei, in which there is more than one spatial orbit, there are no
analogs of Eqs.\ (\ref{eq:phiofrhoi}) and (\ref{eq:j0}) and our
procedure for evaluating $\delta J_0/\delta \rho_I$ won't work. That
quantity is the inverse of a generalized linear response function
for $\rho_I$ at $\omega=0$, and will not usually vanish at $y \neq
x$, as it did in our machinations above.  It be evaluated
numerically, but at a high computational cost. Subleading terms can
still be neglected in this inverse response function, however, if
the only goal is to include the leading corrections to the
mean-field functional near the minimum. That would mean ignoring the
difference between $\delta J_0/\delta \rho_I(x)$ in Eq.\
(\ref{eq:equal}) and the ordinary RPA inverse response function,
which for a given $J_0$ can be calculated in a relatively
straightforward way \cite{Shl75}. Though $\delta {\cal E}_1/\delta
J_0$, the other quantity in Eq.\ (\ref{eq:equal}), might still not
easy, it could be evaluated as already described or (particularly
for real and more complicated functionals) via the relations in
Ref.\ \cite{Eng97}.

If one wanted to avoid even that difficulty, one could make still
more approximations without changing the solution. One could even go
so far as to evaluate $\delta E_1/\delta J_0$  and $\delta
J_0/\delta \rho_I$  at $J_0=0$, i.e.\ at $\rho_I = \langle
\hat{\rho}_I (x)\rangle_{\rm mf} =(mgN^2/4) {\rm cosh}^{-2}(Nmgx/2)$
(the mean-field approximation to the exact value). That would
actually remove $J_0$ from the right hand side of Eq.\
(\ref{eq:equal}), thus changing the self-consistent procedure
represented by Eq.\ (\ref{eq:kseq}) into a one-step formula:
\begin{equation}
\label{eq:reduc} J_0(x) = \int \frac{\delta J_0(y)}{\delta
\rho_I(x))}\Big|_{J_0=0} \times \frac{\delta {\cal E}_1}{\delta
J_0(y)}\Big|_{J_0=0} dy~,
\end{equation}
which, after multiplying both sides by the response function $\delta
\rho_I(z)/ \delta J_0(x)$ (which equals $\delta \rho_I(x)/ \delta
J_0(z)$), and integrating over $x$, implies
\begin{eqnarray}
\label{eq:onestep} \rho_I(x) &=&  \langle \hat{\rho_I}(x)
\rangle_{\rm mf} \big|_{J_0=0}
 + \int
\frac{\delta \rho_I(x)}
{\delta J_0(y)} \Big|_{J_0=0} J_0(y) dy +\cdots \nonumber\\
&=& \left [ \langle
 \hat{\rho}_I(x)
\rangle_{\rm mf}+\frac{\delta {\cal E}_1}{\delta J_0(x)}\right] \Big|_{J_0=0}
+\cdots~.
\end{eqnarray}
(The manipulations serve to obtain the second line from the first, which is just
a Taylor expansion.) Evaluating Eq.\ (\ref{eq:onestep}) would just
be equivalent to calculating the correction to the mean-field
ground-state intrinsic density by summing rings generated by the
attractive delta function and the first-order insertion of an
effective one-body potential (the third term in Eq.\
(\ref{eq:mfeqj})) or two-body interaction (the second line of Eq.\
(\ref{eq:veff})), with $J(y)$ replaced by $\delta(x-y)$ in both
cases because of the functional derivative. That is eminently doable
and would reproduce the second line of Eq.\ (\ref{eq:exactden}), the
expression for the exact ground-state intrinsic density.

In our model, however, no matter where in this chain of
approximations we decided to stop, the intrinsic-density functional
would not be local, i.e., it would not look like a Skyrme
functional. For that, one would seem to need some form of LDA.  The
answer to the question at the end of the introduction, then, is that
we can write the intrinsic density functional in terms of orbitals,
but it's tough, in this model anyway, to make it Skyrme-like.
Nonetheless, the successive approximations do bring an important
point to light: what we really care about is not so much the density
functional for arbitrarily stressed nuclei as it is the ground-state
energy and intrinsic density itself. In fact, since the nucleus is
never subject to a deforming potential, except in nuclear matter
where wave functions are translationally invariant, 
one might argue that that is {\it all} we care about.  It
actually seems a bit strange to apply density-functional theory, the
main feature of which is ``universality'', to a problem in which
there are no external influences to treat universally (though
density functionals do arise naturally without the accompanying
potentials in the density-matrix expansion \cite{Neg70}).   In any
event, a functional that does not exactly match the definition, Eq.\
(\ref{eq:denf}) but does give the correct energy and intrinsic
density when minimized, and has the additional benefit of being
simple, is all we really want for the study of finite nuclei. In the
next section we show that in our model, a simple Skyrme-like
functional satisfies these requirements.

\section{Phenomenological functional and Skyrme interactions\label{sec:ph}}

We want a Skyrme-like intrinsic-density functional with a minimum at
the correct value to next-to-leading order in $1/N$, that is, a
functional that at its minimum reproduces the ${\cal O}(N^3)$ and
${\cal O}(N^2)$ terms (the latter of which is zero) in Eq.\
(\ref{eq:exacte}), and the second line of Eq.\ (\ref{eq:exactden})
when $\rho_I$ is calculated via  Eq.\ (\ref{eq:intrsd}).   We will
reproduce the intrinsic density if the $\phi$ that solves the
Kohn-Sham equation is
\begin{eqnarray}
\label{eq:phimin} \phi(x) &\equiv& \phi_0(x)+\frac{1}{N} \phi_1(x)\\
\phi_0(x)& =& \frac{\sqrt{(N-1)mg}}{2}\left[ {\rm
cosh}^{-1}(\frac{Nmgx}{2c_N})\right] \nonumber\\
\phi_1(x)&=&s\frac{\sqrt{(N-1)mg}}{4} \left[ \frac{3}{2}{\rm
cosh}^{-3}(\frac{Nmgx}{2c_N})- {\rm cosh}^{-1}(\frac{Nmgx}{2c_N})
\right]~, \nonumber
\end{eqnarray}
where $s \equiv 1+m^2g^2N^2  \mu_{\rm gs}/2 = 1 + \pi^2/6$, and the
$c_N$'s appear because of their presence in Eq.\ (\ref{eq:intrsd}).

To express such a functional through the formalism detailed above, we need to
keep the center-of-mass kinetic energy in the Hamiltonian and
use an effective nucleon mass
\begin{equation}
m^*=\frac{m}{1+\alpha/N}~,
\end{equation} where $\alpha$ is a constant to be
determined later\footnote{Though an effective mass can be included
through the use of an intrinsic kinetic-energy density \cite{Bha05},
we will simply alter the nucleon mass here.}.  We denote by
$E_0^*[\rho_I]$ the mean-field functional obtained by changing
$E_0[\rho_I]$ in Eq.\ (\ref{eq:e0}) in that way (both changes affect
the kinetic term), and write the functional through first order in
$1/N$ as
\begin{equation}
\label{eq:myfun}
E[\rho_I]=E_0^*[\rho_I]+E_1[\rho_I]~, \qquad E_1[\rho_I]=-\int \! dx
\left( \frac{\beta
g}{2N}\rho_I^2 +\frac{\gamma}{6m^*N^3}\rho_I^3\right)~,
\end{equation}
where $\beta$ and $\gamma$ are also constants to be determined
later. Eq.\ (\ref{eq:kseq}), with the omission of all but the first
three terms in Eq.\ (\ref{eq:mfeqj}) (which we argued could be
neglected without changing the minimum of the functional), leads to
the explicit Kohn-Sham-like equation
\begin{equation}
\label{eq:mykseq}
- \frac{\phi^{\prime\prime}}{2m^*}-(N-1+\beta)g
|\phi|^2\phi -\frac{\gamma}{2m^*N} |\phi|^4\phi  = \epsilon \phi~.
\end{equation}
Inserting $\phi=\phi_0 + \phi_1/N$ into this equation, writing
$\epsilon \equiv \epsilon_0 + \epsilon_1/N$, and expanding in
inverse powers of $(N-1)$ rather than $N$ (because of the $c_N$'s in
Eq.\ (\ref{eq:phimin})), we find at leading order in $(N-1)^{-1}$
the self-consistent equation
\begin{equation}
\label{eq:phi0} -\frac{1}{2m}\phi_0^{\prime\prime}-(N-1)g|\phi_0|^2
\phi_0 = \epsilon_0 \phi_0~.
\end{equation}
At ${\cal O}(1/[N-1]) \equiv {\cal O}(1/N)$, assuming $\phi$ real,
we have
\begin{equation}
\label{eq:phi1}
- \frac{1}{2m}\phi_1^{\prime\prime}-3(N-1)g \phi_0^2 \phi_1
-\frac{\alpha
}{2m}\phi_0^{\prime\prime} -(N-1)g\beta \phi_0^3 -\frac{\gamma}{2m} \phi_0^5  = \epsilon_0
\phi_1 + \epsilon_1 \phi_0~
\end{equation}

Now, the solution to Eq.\ (\ref{eq:phi0}) is the $\phi_0$ of Eq.\ (\ref{eq:phimin}),
with $\epsilon_0=-mg^2(N-1)^2/8$.  The solution to Eq.\ (\ref{eq:phi1}),
with the appropriate $\phi_0$ and $\epsilon_0$ plugged in,
is the  $\phi_1$ of Eq.\ (\ref{eq:phimin}),
provided $\beta=\alpha-2s$, $\gamma=18s$, and $\epsilon_1=\alpha \epsilon_0$.
The requirement that the energy $E[\rho_I]$ in Eq.\ (\ref{eq:myfun}) have no
${\cal O}(N^2)$ term at the minimum is
satisfied as well if and only if
\begin{equation}
\label{eq:abc}
\alpha = 2(1+\frac{4s}{5})~, \ \ \ \beta=2(1-\frac{s}{5})~, \ \ \ \gamma=18s~.
\end{equation}
Thus, minimizing our functional in Eq.\ (\ref{eq:myfun}) with these
values of the parameters yields the correct energy and intrinsic
density to ${\cal O}(1/N)$.

The functional can be extended to reproduce $\rho_I$ and $E$ at
higher orders in $(1/N)$ with the addition of higher powers of
$\rho_I/N^2$, though finding the corresponding coefficients gets
harder.  These functionals, as discussed, do not converge to the ``exact
one'' for all $\rho_I$, but they do around the ground-state
$\rho_I$. The result suggests that real Skyrme functionals, which may be viewed
as depending on intrinsic semilocal kinetic, spin-orbit, and current densities
in addition to
the ordinary intrinsic density, behave
the same way; we have no reason to think them correct away from the
minimum.


It's worth noting that the construction above is even more
straightforward if we take $\rho_I(x) \equiv N |\phi(x+\bar{x})|^2$
rather than define it as the expectation value of $\hat{\rho_I}$ in
the the Slater determinant constructed from $\phi$, i.e., as in Eq.\
(\ref{eq:intrsd}).  We do the latter only because the mean-field
approximation and, more importantly, \emph{real Skyrme functionals}
work better with that definition \cite{Rei06}.  The success of the
wave-function-based assignment $\rho_I \equiv \langle
\hat{\rho}_I\rangle_{\rm mf}$ raises the
hope that Skyrme functionals can be derived and improved through
mean-field theory with some effective low-momentum interaction, with
the addition of relatively small corrections.  The density-matrix
expansion works along those lines, and deserves more attention. It
remains to be seen whether the formalism presented here --- the
inversion method modified for self-bound systems --- can be happily
married to such an approach.

We have said nothing here about ``symmetry restoration'', a
technique that, particularly in its rotational form, is viewed as a
step beyond the local or semi-local Skyrme-mean-field equations. The
projected-mean-field equations, while more complicated than their
unprojected counterparts, are still equations for single-particle
orbitals, however.  Symmetry-restored mean-field theory is simply a
more accurate starting point for the inversion method than plain
mean-field theory.  Presumably, additional corrections can be
absorbed into those equations, as they were here.  This statement,
of course, says nothing about what the corrected functionals will
look like.

To summarize, we have shown that a version of the Hohenberg-Kohn theorem holds
for the intrinsic density in self-bound system and that, to the extent that a
convergent expansion around mean-field solutions exists for such systems, a
Kohn-Sham-like procedure can be applied to obtain functionals of the intrinsic
density that include all correlations.  Skyrme functionals seem to be
approximations to the exact intrinsic-density functionals but it is quite possible that,
like our
phenomenological functional, they work poorly away from the ground-state
density and energy.  On the other hand, we do not always need them to do more
than that, and a concerted effort to improve even this limited kind of
functional is worthwhile.  The density-matrix expansion, which can clearly be
applied more accurately than it was 35 years ago, may offer improvements and
should be further explored.

Fruitful discussions with A.\ Bhattacharyya, T.\ Duguet, R.J.\
Furnstahl, T.\ Schaefer, and A.\ Schwenk are gratefully acknowledged. This work
was supported in part by the U.S.\ Department of Energy under
Contract DE-FG02-97ER41019.

\end{document}